\documentclass[a4paper]{iopart}
\newcommand{\be}{\begin{equation}}
\newcommand{\ee}{\end{equation}}

\newcommand{\ba}{\begin{array}}
\newcommand{\ea}{\end{array}}
\newcommand{\bea}{\begin{eqnarray}}
\newcommand{\eea}{\end{eqnarray}}
\newcommand{\nn}{\nonumber \\}

\usepackage{graphicx}
\usepackage{iopams, amssymb}
\newcommand{\binom}[2]{\scalebox{.8}{$\Bigl(\begin{array}{@{}c@{}}#1\\#2\end{array}\Bigr)$}}
    
\newtheorem{theorem}{Theorem}

\newtheorem{corollary}{Corollary}
\newtheorem{lemma}{Lemma}

\usepackage{graphicx}
\usepackage{dcolumn}
\usepackage{bm}
\usepackage{cite} 

\begin{document}


\title{General Nth order integrals of the motion}

\author{S. Post$^1$ and P. Winternitz$^2$\footnote{On Sabbatical at Dipartimento di Matematica e Fisica, Universit\`a di Roma Tre, Via della Vasca Navale 84, 00146, Roma, Italy} }%

\address{$^1$ 
Department of Mathematics,
University of Hawai'i at M\=anoa\\
2625 McCarthy Mall, Honolulu (HI) 96822, USA
}%

\address{$^2$
D\'epartement de Math\'ematiques et de Statistique and Centre de Recherches Math\'ematiques,
Universit\'e de Montr\'eal.\\ CP6128, Succursale Centre-Ville, Montr\'eal   (QC) H3C 3J7, Canada 
}%

\date{\today}
\ead{  \mailto{spost@hawaii.edu}, \mailto{wintern@CRM.umontreal.ca}}
\begin{abstract}
 The general form of an integral of motion  that is a polynomial of order
$N$ in the momenta is presented for a Hamiltonian system in two-dimensional
Euclidean space. The classical and the quantum cases are treated
separately, emphasizing both the similarities and the differences between
the two. The main application will be to study Nth order superintegrable
systems that allow  separation of variables in the Hamilton-Jacobi and
Schr\"odinger equations, respectively.
\end{abstract}


\section{Introduction}

The purpose of this article is to provide a framework for systematically studying finite-dimensional integrable and superintegrable systems with integrals of motion that are polynomials of arbitrary finite order, N, in the momenta. In the process, we also establish some basic properties of the integrals of the motion and study some differences between the integrals in classical and quantum mechanics. 

We restrict ourselves to a two-dimensional real Euclidean plane and to Hamiltonians of the form 
\be H=p_1^2+p_2^2+V(x,y).\label{Hgen} \ee
In classical mechanics $p_1$ and $p_2$ are the components of the linear momentum, to which we add for use below the angular momentum 
\[ L_3=xp_2-yp_1.\]
In quantum mechanics, $p_1$ and $p_2$ (as well as $H$ and $L_3$) will be Hermitian operators with 
\be \hat{p}_1=-i\hbar \partial_x, \qquad \hat{p}_2=-i\hbar \partial_y.\label{quantum momenta}\ee

In classical mechanics an Nth order integral of the motion can be written as 
\be X=\sum_{k=0}^N\sum_{j=0}^{N-k} f_{j,k}(x,y)p_1^jp_2^{N-k-j}, \qquad f_{j,k}(x,y)\in \mathbb{R},\label{Xclassbounds}\ee 
or simply
\be X=\sum_{j,k} f_{j,k}(x,y) p_1^{j}p_2^{N-k-j} \label{Xclass},\ee
with $f_{j,k}=0$ for $j<0,\, k<0$ and $k+j>N$. The leading terms (of order $N$) are obtained by restricting the summation to $k=0$. In quantum mechanics we also take the integral of the form (\ref{Xclassbounds}) (or (\ref{Xclass})) but the $p_i$ are operators as in (\ref{quantum momenta})
and we must symmetrize in order for $X$ to be Hermitian (see Section \ref{Quantum Section} below). 

We recall that a Hamiltonian with $n$ degrees of freedom in classical mechanics is integrable (Liouville integrable) if it allows $n$ integrals of motion (including the Hamiltonian) that are well defined functions on phase space, are in involution (Poisson commute) and are functionally independent. The system is superintegrable if it allows more than $n$ integrals that  are functionally independent and  commute with the Hamiltonian. The system is maximally superintegrable\cite{superreview} if it allows $2n-1$ functionally independent, well defined integrals, though at most $n$ of them can be in involution. 

In quantum mechanics the definitions are similar. The integrals are operators in the enveloping algebra of the Heisenberg algebra $H_n\sim \{ x_1, \ldots, x_n, p_1, \ldots, p_n, \hbar \}$ that are either polynomials or convergent power series. A set of integrals is algebraically independent if no Jordan polynomial (formed using only anti-commutators) in the operators vanishes. 

The best known superintegrable systems are the Kepler-Coulomb system\cite{Fock,Bargmann} with the potential $V=\alpha/r$  and the harmonic oscillator\cite{JauchHill1940, MoshSmir} with $V=\alpha r^2$. According to Bertrand's theorem \cite{Bertrand}, the only rotationally invariant potentials in which all bounded trajectories are closed are precisely these two potentials. A theorem proven by Nekhoroshev \cite{nekhoroshev1972action} states that, away from singular points, if a Hamiltonian system in $n$ dimensions is maximally superintegrable ($2n-1$ integrals of motion) then all bounded trajectories are periodic. It follows that there are no other rotationally invariant maximally superintegrable systems in a Euclidean space $E_n$. 

A systematic study of other superintegrable systems started with the construction of all quadratically superintegrable systems in $E_2$ and $E_3$ \cite{WSUF,MSVW1967,FMSUW}. Superintegrable systems with second-order integrals of motion are by now well understood both in spaces of constant curvature  and in more general spaces \cite{KKW, KKMW2003, KKM20041, KKM20042, KKM20051, KKM20061, KKM20062, EVA,  EVAN, Evans1991, rodriguez2002quantum, TempTW}. Second-order superintegrability is related to multi-separability in the Hamilton-Jacobi equations and the Schr\"odinger equation. The superintegrable potentials are the same in classical and quantum mechanics. When commuted amongst each other, the integrals of motion form quadratic algebras \cite{KKMW2003,KKM20042, Zhedanov1992a, Zhedanov1992b, Dask2001, DASK2007, DT2007, VILE}. 

Superintegrable systems involving one third-order and one lower order integral of motion in $E_2$ have been studied more recently \cite{Gravel, GW,  MW2008, TW20101, PopperPostWint2012}. The connection with multiple separation of variables is lost. The quantum potentials are not necessarily the same as the classical ones, i.e. quantum potentials that depend on the Planck constant appear. Some of the quantum potentials obtained involve elliptic functions or Painlev\'e transcendents. The integrals of motion form polynomial algebras and these can be used to calculate energy spectra and wave functions \cite{ marquette2009painleve, marquette2010superintegrability, MW2008}. A relation with supersymmetry has been established \cite{quesne2008exceptional, marquette2009super, marquette2011infinite, marquette2013new}. It was recently shown that infinite families of two-dimensional superintegrable systems exist with integrals of arbitrary order \cite{TTW2009, TTW2010, PW20101, marquette2011infinite, LPW2011, KKM2010JPA, KMPTTWClass, PostRiglioni2014}. Superintegrable systems not allowing separation of variables have been constructed \cite{PW2011, MPY2010, MPT}

The present article is to be viewed in the context of a systematic study of integrable and superintegrable systems with integrals that are  polynomials in the momenta, especially for those of degree higher than two.  Here we concentrate on the properties of one integral of order N in two-dimensional Euclidean space. 

The remainder of the article is organized as follows.  Section 2 is devoted to N-th order integrals of motion in classical
mechanics and includes a derivation of the classical determining equations. The determining equations for quantum integrals are derived in Section
3. They are shown to have the same form
as the classical ones, up to quantum corrections. These corrections are shown to be
polynomials in the square of the Planck constant $\hbar$ and are presented
explicitly. The general formulas are specialized to low order cases N= 2, $\ldots, $5 in Section 4. Different possible quantization procedures are
compared and special cases are considered in Section 5. Conclusions and a
future outlook are presented in Section 6.

\section{Classical Nth order integrals of the motion}
Let us consider the classical Hamiltonian (\ref{Hgen}) and the Nth order integral (\ref{Xclass}) where $f_{j,k}(x,y)$ are real functions. Since $X$ of (\ref{Xclass}) is assumed to be  an integral of the motion, it must Poisson commute with the Hamiltonian
\be \label{XH} \{H, X\}_{PB}=0.\ee
The commutation relation (\ref{XH}) leads directly to a simple but powerful theorem. 

\begin{theorem}
A classical Nth order integral for the Hamiltonian (\ref{Hgen}) has the form 
\be \label{Xclassparity} X=\sum_{\ell=0}^{[\frac{N}{2}]} \sum_{j=0}^{N-2\ell}f_{j,2\ell}p_1^jp_2^{N-j-2\ell},\ee
where  $f_{j,2\ell}$ are real functions that are identically $0$ for $j,\ell<0$ or $j>N-2\ell$, with the following properties: 
\begin{enumerate}
\item The functions $f_{j, 2\ell}$ and the potential $V(x,y)$ satisfy the determining equations
\be\fl  \label{classical deteq} 0=2\bigg(\partial_x f_{j-1,2\ell}+\partial_y f_{j, 2\ell}\bigg)-\bigg((j+1)f_{j+1, 2\ell-2} \partial_x V+(N-2\ell+2-j)f_{j, 2\ell-2}\partial_y V \bigg).\ee
\item As indicated in (\ref{Xclassparity}), all terms in the polynomial $X$ have the same parity.
\item The leading terms in (\ref{Xclassparity})  (of order N obtained for $\ell=0$) are polynomials of order $N$ in the enveloping algebra of the Euclidean Lie algebra $e(2)$ with basis $\{p_1, p_2, L_3\}.$ 
\end{enumerate}
\end{theorem}
{\bf Proof} We calculate the Poisson commutator (\ref{XH}) for H as in (\ref{Hgen}) and $X$ as in (\ref{Xclass})
\bea \fl \{ H, X\}_{PB}&=&\sum_{j,k}-2\frac{\partial f_{j,k}}{\partial x}p_1^{j+1}p_2^{N-k-j}-2\frac{\partial f_{j, k}}{\partial y }p_1^{j}p_2^{N-k-j+1}\nonumber\\\fl  &&+j\frac{\partial V}{\partial x} f_{j, k} p_1^{j-1}p_2^{N-k-j}+(N-k-j)\frac{\partial V}{\partial y} f_{j, k}p_1^jp_2^{N-k-j-1}. \eea
The first two terms are of order $N-k+1$, the second of order $N-k-1$. We shift $j\rightarrow j-1$ in the first term and $j\rightarrow j+1$ in the third to obtain
\bea\fl  0=\sum_{j,k}-2\left(\frac{\partial f_{j-1,k}}{\partial x}+\frac{\partial f_{j, k}}{\partial y }\right)p_1^{j}p_2^{N-k-j+1}\nonumber\\
+\left((j+1)\frac{\partial V}{\partial x} f_{j+1, k}+(N-k-j)\frac{\partial V}{\partial y} f_{j, k}\right)p_1^jp_2^{N-k-j-1}.\label{25} \eea
The terms of order $N+1$ in  the momenta, in  (\ref{25}), correspond to  $k=0$ and are
\be \frac{\partial f_{j-1,0}}{\partial x}+\frac{\partial f_{j,0}}{\partial y} =0. \label{enveloping}\ee
Eq. (\ref{enveloping}) is the condition for the highest order terms of $X$ to Poisson commute with the free Hamiltonian 
$H_0=p_1^2+p_2^2$, thus proving the third statement in Theorem 1. From the form of (\ref{25}), we see that even and odd terms in X are independent. This proves statement 2 of the theorem. Finally, we shift $k\rightarrow k+2$ in the second term of (\ref{25}). The coefficient of $p_1^jp_2^{N-k-j+1}$ (after the shift) must vanish independently for all $(j,k)$ and we obtained the determining equations (\ref{classical deteq}).  

\begin{corollary}
The classical integral (\ref{Xclassparity}) can be rewritten as
\be \fl  \label{XclassparityA} X=\sum_{0\leq m+n\leq N}A_{N-m-n,m,n} L_3^{N-m-n} p_1^{m}p_2^n+ \sum_{\ell=1}^{\lfloor \frac{N}{2}\rfloor} \sum_{j=0}^{N-2\ell}f_{j,2\ell}p_1^jp_2^{N-j-2\ell},\ee
where $A_{N-m-n,m,n}$ are constants. 
\end{corollary}
{\bf Proof:} As noted above, the determining equations for $f_{j,0}$ given by (\ref{enveloping}) do not depend on the potential. The solutions of \eref{enveloping} are 
\be \label{fj0 Nthorder}  f_{j,0}=\sum_{n=0}^{N-j} \sum_{m=0}^{j}\binom{ N-n-m}{j-m}A_{N-n-m,m,n}x^{N-j-n}(-y)^{j-m},\ee
which give the form of the integral (\ref{XclassparityA}). Thus for all N the solutions of \eref{classical deteq} for $\ell=0$ are known in terms of the
$(N+1)(N+2)/2$ constants
$A_{N-n-m,m,n}$ figuring in \eref{XclassparityA}.

Let us add some comments. 
\begin{enumerate}
\item For physical reasons (time reversal invariance) we have assumed that the functions $f_{j,k}(x,y) \in \mathbb{R}$ from the beginning. This is actually no restriction. If $X$ were complex, its real and imaginary parts would Poisson commute with $H$ separately (for $V(x,y)\in \mathbb{R}$) and hence each complex integral would provide two real ones. 
\item The number of determining equations (\ref{classical deteq}) is equal to 
\be\label{numdet} \sum_{\ell=0}^{[\frac{N+1}{2}]}(N-2\ell +2)=\left\{ \ba{ll} \frac14(N+3)^2 & \mbox{N odd}\\
\frac14(N+2)(N+4) & \mbox{ N even.}\ea \right.\ee
\end{enumerate}

For a given potential $V(x,y)$ the equations are linear first-order partial differential equations for the unknowns $f_{j,2\ell}(x,y)$. The number of unknowns is 
\be \label{numcof} \sum_{\ell=0}^{[\frac{N+1}{2}]}(N-2\ell +1)=\left\{ \ba{ll}\frac{(N+1)(N+3)}4 & \mbox{N odd}\\
\frac14(N+2)^2 & \mbox{ N even.}\ea \right.\ee  
As is clear from Corollary 1, the determining equations for $f_{j,0}$ can be solved without knowledge of the potential and the solutions  depend on $(N+1)(N+2)/2$ constants. Thus, $N+1$ of the functions $f_{j,2\ell}$, namely $f_{j,0},$ are known in terms of $(N+1)(N+2)/2$ constants. The remaining system is overdetermined and subject to further compatibility conditions. 
If the potential $V(x,y)$ is not a priori known, then the system (\ref{classical deteq}) becomes nonlinear
and $V(x,y)$ must be determined from the compatibility conditions. We present the first set of compatibility conditions as a corollary. 

\begin{corollary}
If the Hamiltonian (\ref{Hgen}) admits $X$ as an integral then the potential function $V(x,y)$ satisfies the following linear partial differential equation (PDE)
 \be 0=\sum_{j=0}^{N-1} \partial_x^{N-1-j}\partial_y^{j}(-1)^{j}\left[(j+1)f_{j+1, 0} \partial_xV+(N-j)f_{j, 0}\partial_yV\right].\label{linearcomp}\ee
\end{corollary}
{\bf Proof:} This linear PDE is determined by the compatibility conditions for the $\ell=1$ set of determining equations, namely 
\be \label{2ndhighestClassical}2 \left(\partial_xf_{j-1,2}+\partial_yf_{j,2}\right) -\left[(j+1)f_{j+1, 0} \partial_xV+(N-j)f_{j, 0} \partial_yV\right]=0.\ee
Therefore when $X$ is an integral, the functions $f_{j,2}$ exist and satisfy (\ref{2ndhighestClassical}) and so the potential satisfies (\ref{linearcomp}). This PDE depends only on the constants $A_{j,k,\ell}$ in the highest order terms of $X$, \eref{XclassparityA}.

\begin{enumerate}
\setcounter{enumi}{2}

\item For $N$ odd, the lowest-order determining equations are 
\be f_{1,N-1}V_x+f_{0,N-1}V_y=0.\label{lowestclassical}\ee
\end{enumerate}
In particular, for the $N=3$ case, the compatibility conditions of \eref{lowestclassical} with the determining equations  for $f_{j,2}$ lead to nonlinear equations for the potential\cite{GW, TW20101}.

\section{Quantum Nth order integrals of the motion} \label{Quantum Section}
In this section, we shall present a theorem analogous to the classical one of the previous section, but applied instead to  quantum systems. That is, we would like to show that given an Nth order differential operator $X$ that is formally self-adjoint with respect to the Euclidean metric and that commutes with a given Hamiltonian, then the number of independent functions and determining equations are equal to that of the classical case and the determining equations are the same as in the classical case, up to quantum corrections which are polynomial in $\hbar^2. $ Specifically, we have the following theorem.

\begin{theorem}
A quantum Nth order integral for the Hamiltonian (\ref{Hgen}) has the form 
\be \label{Xquantparity} X=\frac12\sum_{\ell=0}^{\lfloor \frac{N}2\rfloor} \sum_{j=0}^{N-2\ell} \lbrace f_{j,2\ell},\hat{p}_1^j\hat{p}_2^{N-2\ell-j}\rbrace,\ee
where $f_{j,2\ell}$ are real functions that are identically $0$ for $j,\ell<0$ or $j>N-2\ell$,
with the following properties: 
\begin{enumerate}
\item  The functions $f_{j, 2\ell}$ and the potential $V(x,y)$ satisfy the determining equations 
\be\label{quant deteq} 0=M_{j,2\ell},\ee
with 
\bea\fl  M_{j,2\ell}&\equiv & 2\left( \partial_xf_{j-1,2\ell}+\partial_yf_{j,2\ell}\right)\nonumber\\
\fl &&-\left((j+1)f_{j+1, 2\ell-2} \partial_xV+(N-2\ell+2-j)f_{j, 2\ell-2} \partial_yV +\hbar^2 Q_{j,2\ell}\right),\nonumber \eea
where $Q_{j,2\ell}$ is a quantum correction term given by 
\bea \label{Quantcorrection}\fl  Q_{j,2\ell}&& \equiv\left(2\partial_x\phi_{j-1,2\ell}+2\partial_y\phi_{j,2\ell} +\partial_x^2\phi_{j,2\ell-1}+\partial_y^2\phi_{j,2\ell-1}\right)\\
\fl &&-\sum_{n=0}^{\ell-2}\sum_{m=0}^{2n+3}(-\hbar^2)^n\binom{j+m}{m}\binom{N-2\ell+2n+4-j-m}{ 2n+3-m}(\partial_x^m\partial_y^{2n+3-m}V)f_{j+m,2\ell-2n-4}\nn
\fl &&-\sum_{n=1}^{2\ell-1}\sum_{m=0}^{n}(-\hbar^2)^{\lfloor(n-1)/2\rfloor}\binom{j+m}{m}\binom{ N-2\ell+n+1+j-m}{ n-m}(\partial_x^m\partial_y^{n-m}V)\phi_{j+m,2\ell-n-1}, \nonumber\eea
where the $\phi_{j,k}$ are defined for $k>0$ as 
\be\fl  \phi_{j,2\ell-\epsilon} =\sum_{b=1}^{\ell}\sum_{a=0}^{2b-\epsilon}\frac{(-\hbar^2)^{b-1}}{2}\binom{j+a}{a}\binom{ N-2\ell+2b-j-a}{2b-\epsilon-a}\partial_x^a\partial_y^{2b-\epsilon-a}f_{j+a, 2\ell-2b} \label{phieven},\ee 
with $\epsilon=0,1.$ In particular $\phi_{j,0} =0$, hence $Q_{j,0} = 0$ so the $\ell=0$ determining are the
same as in the classical case.
\item As indicated in (\ref{Xquantparity}), the symmetrized integral will have terms which are differential operators of the same parity. 
\item The leading terms in (\ref{Xquantparity})  (of order N obtained for $\ell=0$) are polynomials of order $N$ in the enveloping algebra of the Euclidean Lie algebra $e(2)$ with basis $\{\hat{p}_1, \hat{p}_2, \hat{L}_3\}.$ 
\end{enumerate}
\end{theorem}

Notice that the  parity  constraint on the integral \eref{Xquantparity} reduces the number of possible functional coefficients by about half. Indeed, if the integral is expanded out with the derivatives moved to the left then the integral would be of  form 
\bea \fl X&=&\sum_{\ell, j}\left(f_{j,2\ell} -\hbar^2\phi_{j,2\ell}\right)\hat{p}_1^{j}\hat{p}_2^{N-2\ell-j}-i\hbar\sum_{\ell, j}\phi_{j, 2\ell-1}\hat{p}_1^j\hat{p}_2^{N-2\ell+1-j}.\label{Xphi} \eea 
Thus, for a general, self-adjoint $Nth$-order integral that commutes with $H$, approximately half of the coefficient functions depend only on derivatives of the functions $f_{j,2\ell},$ these are the $\phi_{j,2\ell-1}$. In general, the functions $\phi_{j,k}$ \eref{phieven}  are polynomial in $\hbar^2$.

We begin the proofs by showing that, modulo lower order integrals of motion, $X$ can be taken to be self-adjoint. 
\begin{lemma}\label{selfadjoint}
Given $X$ an Nth order differential operator  that commutes with a self-adjoint Hamiltonian $H$, then $X$ can be assumed to be self-adjoint.
\end{lemma}
Proof. We can always write $X=\frac{X+X^\dagger}{2}+\frac{X-X^\dagger}{2}.$ Using this, we obtain
\[ 0=[\frac{X+X^\dagger}{2},H]+[\frac{X-X^\dagger}{2},H]\] 
and its Hermitian conjugate 
\[ 0=-[\frac{X+X^\dagger}{2},H]+[\frac{X-X^\dagger}{2},H]\]
which together show that the self-adjoint and skew-adjoint part of the operator $X$ must simultaneously commute with $H.$ Since, $\frac{X-X^\dagger}{2}$ commutes with $H$ so will $i\frac{X-X^\dagger}{2}$ and hence it is possible to assume, without loss of generality,  that the integral is already in self-adjoint form. Q.E.D. 

Next, we show that a commutator of the form $[f, \partial_x^j\partial_y^{1+k-j}]$ can be written as a sum of lower order anti-commutators which differ in degree by 2.  We assume for convenience that all of the functions are smooth but of course this could be reduced to require only the number of necessary derivatives, in this case $k+1.$
\begin{lemma}
Given a real $f\in C^{\infty}(\mathbb{R}^2),$ there exist real functions, $f_{m,n}\in C^{\infty}(\mathbb{R}^2),$  that satisfy the equality
\be \label{lemma2} [f,\partial_x^j\partial_y^{1+k-j}] =\sum_{n=0}^{k}\sum_{m=\max\{0, n-k+j\}}^{\min\{j, n+1\}}\lbrace f_{m,n}, \partial_x^{j-m}\partial_y^{k-n-j+m}\rbrace \ee
 and furthermore the functions with odd second index,  $f_{m, 2\ell+1},$ are identically $0.$\end{lemma}
 Proof. We define the $f_{m,n}$ inductively on $n.$ Notice that the summation indices  on the right-hand side of the identity (\ref{lemma2}) lie within the following regime 
 \be \label{indices regimes} \Lambda=\{ (m,n) | 0\leq j-m\leq j , \, 0\leq k-n-j+m\leq k-j+1\}.\ee Furthermore, the index $n$ determines the total degree of the monomial $\partial_x^{j-m}\partial_y^{k-n-(j-m)}$.  We define the identity operator to be $\partial_x^0\partial_y^0=Id.$
 
Consider the quantity 
\be U= [f,\partial_x^j\partial_y^{1+k-j}] -\sum_{n,m\in \Lambda}\lbrace f_{m,n}, \partial_x^{j-m}\partial_y^{k-j-n+m}\rbrace.\ee
Expanding out $U$ gives
\[U= -2\sum_{n,m \in \Lambda} U_{n,m} \partial_x^{j-m}\partial_y^{k-n-j+m},\]
with 
\bea \fl U_{n,m} &&\equiv\bigg( f_{m,n}+\frac12\binom{ j}{m}\binom{k-j+1}{ 1+n-m}\partial_x^m\partial_y^{1+n-m}f\\
\fl &&+\frac12  \sum_{b=0}^{n-1}\sum_{a\in \sigma}{ \binom{ j-m+a}{ a}\binom{k-n-j+m+1+b-a}{ 1+b-a}}\partial_x^a\partial_y^{1+b-a}f_{m-a,n-1-b}\bigg), \nonumber \eea
where $\sigma=\{\max\{0, b-n+m\}\ldots \min\{m, k-j-n+b+1+m\}\}.$
 Hence for a fixed $m,n$ we have a recurrence relation on the $f_{m,n}$'s which will make each $U_{m,n}=0$ this relation is a recurrence on $n$ given by 
 \bea \fl f_{m,n}&=&-\frac12 \binom{j}{m} \binom{k-j+1}{ 1+n-m}\partial_x^m\partial_y^{1+n-m}f\\
 \fl &-&\frac12 \sum_{b=0}^{n-1}\sum_{a\in \sigma }{\binom{ j-m+a}{ a}\binom{k-n-j+m+1+b-a}{1+b-a}}\partial_x^a\partial_y^{1+b-a}f_{m-a,n-1-b}. \nonumber \eea 
By induction on $n$, the functions $f_{m,n}$ exist and are real. The first two functions ($n=0$) are 
\[f_{0,0}=-\frac{1}{2}\partial_yf, \qquad f_{1,0}=-\frac12 \partial_x f, \qquad j>0,\quad  1+k-j>0.\]
If $j=0$ or $1+k-j=0$, then either $f_{1,0}=0$ or $f_{0,0}=0$, respectively. The summation over $a$ and $b$ ensures that $(m-a, n-1-b)\in \Lambda$ and so  the functions are well-defined recursively on n for all $m,n \in \Lambda$. 

Finally, we see that if $k$ is even then $k+1$ will be odd and so the operator  $[f, \partial_x^j\partial_y^{1+k-j}]$ will be self-adjoint which gives the requirements that the sum 
 $$\sum_{n=0}^{k}\sum_{m=0}^{j}\lbrace f_{m,n}, \partial_x^{j-m}\partial_y^{k-n-j+m}\rbrace$$ contains only even terms in $k-n$ and hence only terms with $n=2\ell$ can be non-zero. Similarly, for $k$ odd, $k+1$ is even and so both sides must be skew-adjoint and hence $k-n$ must be odd which again gives the requirement that  $n=2\ell.$ Q.E.D. 
 
 Next, we use the previous  lemma  to show that any  self-adjoint operator $X$ can be put into symmetric form so that the functional coefficients are real. 

\begin{lemma}\label{sym}
Given a general, self-adjoint  $N$th order differential operator, $X$, there exist real functions $f_{j,k}$ such that 
\be X=\frac12\sum_k\sum_j \lbrace f_{j,k}, \hat{p}_1^j\hat{p}_2^{N-k-j}\rbrace\ee
\end{lemma}
Proof. Given a self-adjoint Nth-order differential operator $X,$ we can always move all of the functional coefficients to the left to obtain 
\be\label{naiveL'} X=\sum_k\sum_j F_{j,k}(-i\hbar)^{N-k}\partial_x^j\partial_y^{N-k-j}, \ee
\[ F_{j,k}\equiv 0, \, \forall \ j<0, k<0, j>k, k>N, \]
where the functions $F_{j,k}$ are possibly complex; we write $F_{j,k}=F_{j,k}^R+iF_{j,k}^I.$  Since $X$ is self-adjoint, it can be expressed as 
\be\fl  X=\frac 12(X+X^\dagger)=\frac12\sum_k\sum_j\left( \lbrace F^R_{j,k},\partial_x^j\partial_y^{N-k-j}\rbrace+i[ F^I_{j,k},\partial_x^j\partial_y^{N-k-j}]  \right)(-i\hbar)^{N-k}. \ee
Next, we use the previous lemma to show that for a given $j,k$ there exists real functions, call them $g^{j,k}_{m,n}$ such that 
\bea \fl  i [ F^I_{j,k}\partial_x^j\partial_y^{N-k-j}](-i\hbar)^{N-k}=\sum_{n=0}^{\lfloor\frac{N-k-1}{2}\rfloor}\sum_{m=0}^{j} \lbrace (-1)^{n}\hbar^{2n+1}g^{j,k}_{m,2n}, \hat{p}_1^{j-m}\hat{p}_2^{N-k-1-2n-j+m}\rbrace.\nonumber\eea
For simplicity, we defined $g_{m,2n}^{j,k}\equiv 0$ whenever $(m, 2n)\notin \Lambda$ (\ref{indices regimes}). 
Thus, there exist real  functions $f_{j,k}$ such that $X$ can be written as 
\be X=\frac12\sum_k\sum_j \lbrace f_{j,k}, \hat{p}_1^j\hat{p}_2^{N-k-j}\rbrace.\ee
where 
\[ f_{j,k}=F_{j,k}^R+\sum_{n=0}^{[(k-1)/2]}\sum_{m=0}^{j}(-1)^n\hbar^{2n+1} g_{m,2n}^{j+n,k-1-2n}\] 
Q.E.D.\\

{\bf Proof of Theorem 2} First, we prove that, modulo lower-order integrals of the motion,  $X$ contains only even or odd terms. 
 By Lemma \ref{selfadjoint}, modulo lower order terms, $X$ can be taken to be self-adjoint. Then, by Lemma \ref{sym} we know that $X$ can be written in the symmetrized form 
\bea X&=&X^{e}+X^{o}\nn
X^{e}&=&\frac12 \sum_{\ell=0}^{[\frac{N}2]}\sum_{j=0}^{N-2\ell} (-i\hbar)^{N-2\ell}\lbrace f_{j,2\ell}, \partial_x^j\partial_y^{N-2\ell-j}\rbrace\nn
X^{o}&=&\frac12 \sum_{\ell=0}^{[\frac{N-1}2]}\sum_{j=0}^{N-2\ell-1} (-i\hbar)^{N-2\ell-1}\lbrace f_{j,2\ell+1}, \partial_x^j\partial_y^{N-2\ell-j-1}\rbrace \eea
where the $X^{e}$ and $X^{o}$ have the opposite parity under time reversal (complex conjugation). Thus, since $H$ is completely real, we know that $X^{e}$ and $X^{o}$ must independently commute with $H$ and so, modulo the lower-order integral  $X^{o},$ $X$ can be written as 
\be\label{Lfinal}  X=\frac12 \sum_{\ell=0}^{[\frac{N}2]}\sum_{j=0}^{N-2\ell} (-i\hbar)^{N-2\ell}\lbrace f_{j,2\ell}, \partial_x^j\partial_y^{N-2\ell-j}\rbrace. \ee 

Next, we consider the determining equations for  the functions $f_{j,k}$ and the potential $V$ making $X$ a quantum integral.  Take $X$ and $H$ as above both self-adjoint and write 
\be [X,H]\equiv \sum_{k=0}^{1+N}\sum_{j=0}^{1+N-k}M_{j,k}\partial_x^{j}\partial_y^{1+N-k-j}(-i\hbar)^{2+N-k}.\ee
We will show that the terms $M_{j, 2\ell+1}$ can be written as differential consequences of  the $M_{j, 2\ell}$ and that the requirements $0=M_{j, 2\ell}$ are given by \eref{quant deteq},  which are the determining equations for the system. 
We begin by showing that  the terms $M_{j, 2\ell+1}$ can be written as differential consequences of  the $M_{j, 2\ell}.$ We know that $[X,H]$ is skew-adjoint and so we have the requirement that 
\be \label{symM}\fl  \sum_{k=0}^{1+N} \sum_{j=0}^{1+N-k}\left(M_{j,k}\partial_x^{j}\partial_y^{1+N-k-j}(-i\hbar)^{1+N-k}\right)^\dagger+M_{j,k}\partial_x^{j}\partial_y^{1+N-k-j}(-i\hbar)^{1+N-k}=0.\ee
Now, from (\ref{Lfinal}) we know that $(-i)^NX$ will be even with respect to complex conjugation (time reversal) and so we can see that  $[(-i^N)X,H]$ will be real. Thus, since we have the equality
\[ [(-i^N)X,H]=\sum_{k=0}^{1+N}\sum_{j=0}^{1+N-k}(i)^N M_{j,k}\partial_x^{j}\partial_y^{1+N-k-j}(-i\hbar)^{2+N-k},\]
we see that both side will be real differential operators which implies that the even terms,  $M_{j,2\ell},$ are completely imaginary and the odd terms, $M_{j, 2\ell+1},$ are real. 
Hence, we can compute (\ref{symM}) 
\bea\fl  0&=&\sum_{k=0}^{1+N} \sum_{j=0}^{1+N-k}\left(M_{j,k}\partial_x^{j}\partial_y^{1+N-k-j}(-i\hbar)^{2+N-k}\right)^\dagger+M_{j,k}\partial_x^{j}\partial_y^{1+N-k-j}(-i\hbar)^{2+N-k}\nn
\label{oddterm}\fl  &=&\sum_{j,\ell} [M_{j,2\ell},\partial_x^{j}\partial_y^{1+N-2\ell -j}(-i\hbar)^{2+N-2\ell}] +\lbrace M_{j, 2\ell +1}, \partial_x^{j}\partial_y^{N-2\ell -j}(-i\hbar)^{1+N-2\ell}\rbrace .\eea
The coefficient of $\partial_x^{j}\partial_y^{N-2\ell -j}(-i\hbar)^{1+N-2\ell}$ in \eref{oddterm}
gives
\bea\label{oddrecurrsion}\fl  0&=&M_{j,2\ell+1}\nn\fl 
&+&i\hbar\sum_{n=0}^{\ell}\sum_{m=0}^{2n+1}\frac{(-\hbar^2)^{n}}2\binom{j+m}{m}\binom{ N-2\ell+2n+1-j+m}{ 2n+1-m}\partial_x^{m}\partial_y^{2n+1-m}M_{j+m, 2\ell-2n}\nn\fl 
&&-\hbar^2 \sum_{n=0}^{\ell-1}\sum_{m=0}^{2n+2}\frac{(-\hbar^2)^n}{2}\binom{j+m}{m}\binom{N-2\ell +2n+2-j+m}{ 2n+2-m}\partial_x^m\partial_y^{2n+2-m}M_{j+m,2\ell-2n-1}.\nonumber
\eea
 For the case, $\ell=0$ we obtain 
 \[ M_{j,1}=\frac{-i\hbar}{2}\left( (j+1)\partial_xM_{j+1,0}+(N-j+1)\partial_yM_{j,0}\right) \]
and so by induction on $\ell,$ we can see that the odd terms $M_{j, 2\ell+1}$ are linear combinations of derivatives of the $M_{j, 2\ell}.$    

The quantity $M_{j,2\ell}$ can then be directly computed using the expansion of $X$, as in (\ref{Xphi}) with the functions $\phi_{j, k}$ given as in (\ref{phieven}), to obtain \eref{quant deteq}. Notice that this equation agrees with (\ref{classical deteq}) modulo terms which are polynomial in $\hbar^2$ and which vanish in the classical limit ($\hbar\rightarrow 0$). 

Finally, as discussed above, if the requirements $M_{j,2\ell}=0$ and their compatibility equations are satisfied then so too will be $M_{j,2\ell+1}=0$ and so the requirement $[H,X]=0$ will also be satisfied. 

To finish the proof of Theorem 2, we show that the highest order terms of $X$ are in the enveloping algebra of the Lie algebra generated by $\hat{p}_1, \hat{p}_2$ and $\hat{L}_3\equiv y\hat{p}_1-x\hat{p}_2.$
The determining equations for the $f_{j,0}$ are the same as in the classical case \eref{enveloping}  and hence their solutions are the same \eref{fj0 Nthorder}. Thus, if such a symmetry operator exists, then the highest order terms agree with those of an Nth-order operator in the enveloping algebra of the Lie algebra generated by $\hat{p}_1, \hat{p}_2$ and $\hat{L}_3$ and so it is always possible to express the highest order terms as operators in the enveloping algebra, with suitable modification of the lower order terms.  Q. E. D. \\

Just as in the classical case, the highest-order determining equations \eref{enveloping} can be solved directly and the functions $f_{j,0}$ are given by \eref{fj0 Nthorder}. However, as will be discussed in Section 4 below, the choice of symmetrization of the leading order terms will lead to $\hbar^2$-dependent correction terms in the lower-order functions.  The quantum version of Corollary 2 still holds and, remarkably, the linear compatibility condition is the same. This does not imply that the quantum and classical potentials are
necessarily the same since further (nonlinear) compatibility conditions
exist. They are in general different in the two cases. 
\begin{corollary}
If the quantum Hamiltonian (\ref{Hgen}) admits $X$ as an integral then the potential function $V(x,y)$ satisfies the same  linear PDE as in the classical case, namely \eref{linearcomp}.
\end{corollary}
{\bf Proof:} The $\ell=1$ set of determining equations are 
\be \label{2ndhighest}M_{j,2}=0.\ee
with 
\[\fl M_{j,2}=2 \left(\partial_xf_{j-1,2}+\partial_yf_{j,2}\right) -\left[(j+1)f_{j+1, 0}\partial_xV+(N-j)f_{j, 0} \partial_yV +\hbar^2 Q_{j,2}\right],\]
with quantum correction term
\[Q_{j,2}=2\partial_x\phi_{j-1,2}+2\partial_y\phi_{j,2} +\partial_x^2\phi_{j,1}+\partial_y^2\phi_{j,1}.\]
The functions $\phi_{j,k}$, coming from expanding out the highest order terms are 
\[\phi_{j,1} =\frac{j+1}2 \partial_x f_{j+1,0}+\frac{N-j}{2}\partial_y f_{j, 0},\]
\[ \phi_{j,2} =\sum_{a=0}^{2}\frac{1}{2}\left(\ba{c}j+a\\a\ea\right)\left(\ba{c} N-j-a\\ 2-a\ea \right)\partial_x^a\partial_y^{2-a}f_{j+a, 0} . \]
The linear compatibility condition of \eref{2ndhighest} is obtained from 
\bea \label{linearcompatbilityalmost} \fl   0&=&\sum_{j=0}^{N-1} \partial_y^{N-1-j}\partial_x^{j}(-1)^{j-1} M_{j,2}\\
\fl &=&\sum_{j=0}^{N-1} \partial_x^{N-1-j}\partial_y^{j}(-1)^{j}\left[(j+1)\partial_xVf_{j+1, 0}+(N-j)\partial_yVf_{j, 0}+\hbar^2 Q_{j,2}\right] \nonumber.\eea
The coefficient of $\hbar^2$ in this equation vanish,  as the relevant terms are zero, i.e. 
\[ \partial_x^{N-1-j}\partial_y^{j}Q_{j,2}=0,\]
 since the functions $f_{j,0}$ are polynomials of total degree at most $N$, see \eref{fj0 Nthorder}. Thus, the potential satisfies the  linear compatibility condition \eref{linearcomp}, as in the classical case.

In fact, this result can be obtain directly by considering instead the following form of the integral. Consider a general, homogeneous polynomial of degree $N$ in the operators $\hat{p}_1, \hat{p}_2$ and $\hat{L}_3$ that is self-adjoint, call it $P_N(\hat{p}_1, \hat{p}_2, \hat{L}_3)$. By Lemma 2 above, this 
operator can be expanded out as 
\[P_N(\hat{p}_1, \hat{p}_2, \hat{L}_3)=\frac12\{f_{j,0}, \hat{p}_1^j\hat{p}_2^{N-j}\}+\sum_{\ell=1}^{\lfloor N/2 \rfloor} \frac12 \{ \psi_{j, 2\ell}, \hat{p}_1^{j}\hat{p}_2^{N-2\ell-j}\}.\]
 Thus, the operator $X$ given in (\ref{Xquantparity}) can instead be expressed as 
\be \label{Xenvelop} X= P_N(\hat{p}_1, \hat{p}_2, \hat{L}_3)+\sum_{\ell=1}^{\lfloor N/2 \rfloor} \frac12 \{ \tilde{f}_{j, 2\ell}, \hat{p}_1^{j}\hat{p}_2^{N-2\ell-j}\}.\ee
In practice $P_N(\hat{p}_1, \hat{p}_2, \hat{M})$ is often chosen as 
\be \label{standardsym} P_N(\hat{p}_1, \hat{p}_2, \hat{L}_3)=\frac12\sum_{m,n}A_{N-m-n,m,n}\{\hat{p}_1^m\hat{p}_2^n, \hat{L}_3^{N-m-n}\},\ee
although this choice is not necessary for what follows. Different possible choices of symmetrization will be discussed in Section 4. Let us now consider the determining equations for the $\tilde{f}_{j, 2\ell}.$ Recall the $(j,2\ell)$ determining equation is obtained from the coefficient of $\hat{p}_1^j\hat{p}_2^{N-2\ell+1-j}$ in  $[X,H].$ Expanding $[X,H]$ gives 
\be \label{XtH} [X,H]=[P_N, H_0]+[P_N, V]+\left[ \sum_{\ell=1}^{\lfloor N/2 \rfloor}\sum_{j=0}^{N-2\ell} \frac12 \{ \tilde{f}_{j, 2\ell}, \hat{p}_1^{j}\hat{p}_2^{N-2\ell-j}\}, H\right].\ee
By definition, the first commutator is 0. The third commutator will give exactly the determining equations (\ref{quant deteq}) with $f_{j, 2\ell}$ replaced by $\tilde{f}_{j,2\ell}$ except that there will be no $f_{j,0}$ terms in the $\phi_{j,k}$'s. To be precise, expanding out all the terms of $X$ except $P_N$ gives  
\bea\fl   \sum_{\ell=1}^{\lfloor N/2 \rfloor}\sum_{j=0}^{N-2\ell} \frac12 \{ \tilde{f}_{j, 2\ell}, \hat{p}_1^{j}\hat{p}_2^{N-2\ell-j}\}&=&\sum_{\ell=1}^{\lfloor N/2 \rfloor} \sum_{j=0}^{N-2\ell}\left(\tilde{f}_{j, 2\ell}-\hbar^2\tilde{\phi}_{j,2\ell}\right)\hat{p}^j\hat{p}^{N-2\ell-j}\nonumber\\&&-i\hbar\sum_{\ell=1}^{\lfloor N/2 \rfloor} \sum_{j=0}^{N-2\ell-1}\tilde{\phi}_{j,2\ell-1}p_{1}^jp_2^{N-2\ell+1-j},\eea
where 
\[\tilde{\phi}_{j,2\ell-\epsilon}=\phi_{j,2\ell-\epsilon}-\sum_{a=0}^{2\ell-\epsilon}(-\hbar^2)^{\ell-1}\binom{j+a}{a}\binom{N-j-a}{2\ell-\epsilon-a} \partial_x^a\partial_y^{2\ell-\epsilon-a}f_{j+a,0}.\]
Therefore, the coefficient of $\hat{p}_{1}^j\hat{p}_{2}^{N-2\ell-j+1}$ coming from the third commutator in (\ref{XtH}) are  $-i\hbar M_{j,2\ell}$ (\ref{quant deteq}) with $f_{j,k}$ replaced by $\tilde{f}_{j,k}$ and $\phi_{j,k}$ replaced by $\tilde{\phi}_{j,k}$, call these $\tilde{M}_{j,2\ell}$. 
To finish the determining equations, we would need to compute the coefficient of $\hat{p}_1^j\hat{p}_2^{N-2\ell+1-j}$ in $[P_N, V]$. These equations would clearly depend on the choice of symmetrization, discussed later, and even in the case (\ref{standardsym}) are not particularly illuminating.

More interesting to consider is the dependence on $\hbar$ of the determining equations for $\tilde{f}_{j, 2\ell}$.  Notice that the highest order terms in $\hbar$ are absent from the $\tilde{\phi}_{j, 2\ell-\epsilon}$'s and so, whereas in general the functions $\phi_{k, 2\ell-\epsilon}$ have terms of order $\ell $ in $\hbar^2$, the functions   $\tilde{\phi}_{k, 2\ell-\epsilon}$ will have terms of only $\ell-1$. Therefore, while the functions $M_{j, 2\ell}$ and hence the determining equations for $f_{j,2\ell}$ have at most order $\ell$ in $\hbar^2$, the functions $\tilde{M}_{j,2\ell}$ depend at most $\ell-1$. As mentioned above the determining equations for $\tilde{f}_{j,2\ell}$ coming from the coefficient of $\hat{p}_1^{j}\hat{p}_2^{N-2\ell+1-j}$ will be $-i\hbar \tilde{M}_{j,2\ell}$ plus the coefficients coming from $[P_N, V].$ These coefficients will be at most order $\ell-1$ and so will not add to the degree of $\hbar^2$ in the determining equations for the adjusted functions $\tilde{f}_{j,2\ell}$, which will be at most order $\ell-1.$  In particular the second set of equations, those for $\tilde{f}_{j, 2}$, will have no quantum correction terms at all and will be the same as in the classical case. This is in agreement with the previous corollary since the compatibility conditions are independent of the functions $f_{j,2}$ or $\tilde{f}_{j,2}.$

\section{Special Cases: N from 2 to 5}

\subsection{Case $N=2$}
This is the most well-known and researched case because of its connection with separation of variables. However, it proves illustrative for future examples. After expanding, the integral of motion \eref{Xquantparity} becomes
\bea X&=&\sum_{j=0}^{2}  f_{j,0}\hat{p}_1^{j}\hat{p}_2^{2-j}-i\hbar\sum_{j=0}^{1}\phi_{j, 1}\hat{p}_1^j\hat{p}_2^{1-j}+f_{0,2}-\hbar^2\phi_{0,2},\eea 
with
\bea &\phi_{0,1}=\partial_y f_{0,0}+\frac12\partial_x f_{1,0},  &\phi_{1,1}=\partial_x f_{2,0}+\frac12  \partial_yf_{1,0}\nonumber\\
& \phi_{0,2}=\frac12\sum_{a=0}^{2}\partial_x^a\partial_y^{2-a}f_{a,0},\nonumber\eea
 in agreement with \eref{phieven}. 
Notice that these all depend on the highest-order terms $f_{j,0}$ which lie in the enveloping algebra of $E_2$ and are given by \eref{fj0 Nthorder}. 
From these solutions we can immediately see that any third-order derivative of a function $f_{j,0}$ is identically zero. 

The  final determining equations are $M_{j,2}=0$ with
\bea M_{j,2}&=&2\partial_xf_{j-1,2}+2\partial_yf_{j,2}-(j+1)f_{j+1, 0} \partial_x V-(2-j)f_{j, 0}\partial_yV\nn
&&-\hbar^2\left(2\partial_x\phi_{j-1,2}+2\partial_y\phi_{j,2} +\partial_x^2\phi_{j,1}+\partial_y^2\phi_{j,1}\right).\eea
Note that the quantum correction term (the second line) depends on third derivatives of the $f_{j,0}$ and hence vanish on solutions \eref{fj0 Nthorder}. Thus, the determining equations for second-order integrals of the motion are equivalent in both the classical and quantum cases. They reduce to 
\bea M_{0,2}&=&2\partial_y f_{0,2}-f_{1,0}\partial_xV-2f_{0,0}\partial_y V=0\\
M_{1,2}&=& 2\partial_x f_{0,2} -2f_{2,0}\partial_x V-f_{1,0}\partial_y V=0.\eea

\subsection{Case $N=3$}
We now turn to the case $N=3.$ This case was investigated in \cite{Gravel, GW}, where the determining equations were given. Here we obtain the same results, presented in the notation of this paper. 

As for all $N$, the $\phi_{j,0}$ are identically 0. There are thus  essentially two families of the $\phi's:$ those that depend only on $f_{j,0}$ 
\bea \phi_{j,1}=\sum_{a=0}^{1} \frac{1}{2}\left(\ba{c}j+a\\a\ea\right)\left(\ba{c} 3-j-a\\ 1-a\ea \right)\partial_x^a\partial_y^{1-a}f_{j+a, 0}\\
\phi_{j,2}=\sum_{a=0}^{2}\frac{1}{2}\left(\ba{c}j+a\\a\ea\right)\left(\ba{c} 3-j-a\\ 2-a\ea \right)\partial_x^a\partial_y^{2-a}f_{j+a,0}.\eea
and one that also depends on  $f_{j,2}$,
\bea \phi_{0,3}=\sum_{b=1}^{2}\sum_{a=0}^{2b-1} \frac{(-\hbar^2)^{b-1}}{2}\partial_x^a\partial_y^{2b-1-a}f_{a, 4-2b}.\eea
Let us now consider the quantum correction terms. As always we have $Q_{j,0} =0$ \eref{phieven}. The determining equations $M_{j,0}=0$  are given by \eref{enveloping} with solutions \eref{fj0 Nthorder}.
The next set of quantum corrections are 
\bea Q_{j,2}& =&\left(2\partial_x\phi_{j-1,2}+2\partial_y\phi_{j,2} +\partial_x^2\phi_{j,1}+\partial_y^2\phi_{j,1}\right).\eea
While it is not immediately clear that these vanish on solutions of (\ref{enveloping}), one need only observe that each of these terms involve third-order derivatives of the $f_{j,0}$ all of which vanish except for
\be\partial_x^3f_{0,0}=-\partial_x^2\partial_yf_{1,0}=\partial_x\partial_y^2f_{2,0}=-\partial_y^3f_{3,0}=6A_{300}.\label{highest3rdderive}\ee
From these, it is straightforward to compute the only non-trivial terms 
\be \partial_x\phi_{j-1,2}=\partial_x^2\phi_{j,1}=\left\{\ba{lr} 0& j=0\\
-6A_{300}& j=1\ea\right. ,\nonumber\ee
\be \nonumber \partial_y\phi_{j,2}=\partial_y^2\phi_{j,1}=\left\{\ba{lr} 0& j=0\\
6A_{300}& j=1\ea\right. .\ee
Thus,  $Q_{j,2}=0$ on solutions of (\ref{enveloping}) and so the next set of determining equations are again independent of $\hbar$ and given by 
\bea \label{f02} \partial_yf_{0,2}&=&\frac12 f_{1,0}\partial_xV+\frac32f_{0,0}\partial_yV\\
\partial_xf_{0,2}+\partial_yf_{1,2}&=& f_{2,0}\partial_xV+f_{1,0}\partial_yV\\
\partial_xf_{1,2}&=& \frac32 f_{3,0}\partial_xV+\frac12f_{2,0}\partial_yV.\label{f12} \eea
To compare these equations to those of Gravel\cite{Gravel}, the potential is multiplied by a factor of 2 (to make up for the choice  of a factor of 1 in the kinetic energy term in \eref{Hgen}) and the functions $f_{j,0}$ are related to $f_k$ via  $f_{3,0}=f_1$, $f_{2,0}=f_2$, $f_{1,0}=f_3$ and $f_{0,0}=f_4.$ The other set of functions are related via $g_{1}=f_{1,2}-5\hbar^2yA_{300}+\frac32\hbar^2A_{210}$ and $g_{2}=f_{0,2}+5\hbar^2xA_{300}+\frac32\hbar^2A_{201}.$ The quantum correction terms are related to the choice of symmetrization, as discussed in Section 5.

Turning now to the final determining equation. The quantum correction term is
\bea
\fl Q_{0,4}&=&\left( \partial_x^2\phi_{0,3}+\partial_y^2\phi_{0,3}\right)\nonumber\\
\fl &&-\sum_{m=0}^{3}(\partial_x^m\partial_y^{3-m}V)f_{m,0}-\sum_{m=0}^{2}(\partial_x^m\partial_y^{2-m}V)\phi_{m,1}-\sum_{m=0}^{1}(\partial_x^m\partial_y^{1-m}V)\phi_{m,2}\nonumber.\eea 
Note that from this expression it appears that  $Q_{0,4}$ has a term depending on $\hbar^2$ (from $\phi_{0,3}$) which would lead to an $\hbar^4$ term to the final determining equation $M_{0,4}$. However, if we inspect this term, we can see that it contains only fifth-order derivatives of the functions $f_{j,0}$ and so will vanish, see (\ref{fj0 Nthorder}). This  
leads to the simplification
\bea \fl 0&=&-f_{1,2}\partial_xV-f_{0,2}\partial_yV\nn
       \fl    &&-\hbar^2\left(-\frac14\sum_{m=0}^{3}(\partial_x^m\partial_y^{3-m}V)f_{m,0}-\frac12(\partial_x\partial_yf_{2,0})\partial_xV-\frac12\left(\partial_x\partial_yf_{1,0}\right)\partial_yV .\right)\eea
Note that this equation is linear and homogeneous in derivative of the potential $V$. 

Summarizing the results from this section, there are three families of determining equations. The first ensures that the leading order terms are in the enveloping algebra. This set as well as the second set of equations are  the same in the classical and quantum case. The final set of equations (in this case one equation) does have a quantum correction term, linear in $\hbar^2$ which is also linear and homogeneous in the derivatives of the potential V.  
\subsection{Case $N=4$}

The structure of the fourth- order integrals is similar to that for $N=3,$
except that the number of equations in each set will be different. These
determining equations were obtained in Ref.\cite{PW2011}, for completeness we present them here. 
There are 4 equations for the $f_{j,2}$ functions.
\bea \label{f023}
 \partial_yf_{0,2}&=&\frac12 f_{1,0}\partial_xV+2 f_{0,0}\partial_yV +\hbar^2( 6yA_{400}-\frac32A_{310}) \\
\partial_xf_{0,2}+\partial_yf_{1,2}&=& f_{2,0}\partial_xV+\frac{3}{2} f_{1,0}\partial_yV+\hbar^2( 6xA_{400}+\frac32A_{301})\\
\partial_xf_{1,2}+\partial_yf_{2,2}&=& \frac{3}{2}f_{3,0}\partial_xV+ f_{2,0}\partial_yV +\hbar^2( 6yA_{400}-\frac32A_{310})\\
\partial_x f_{2,2}&=&2f_{4,0}\partial_xV+ \frac{1}{2}f_{3,0}\partial_yV +\hbar^2( 6xA_{400}+\frac32A_{301}).\label{f323} \eea
Notice that, unlike the case $N=3$,  the quantum corrections $Q_{j,2}$ are not identically 0. However, the equivalent computations obtained in \cite{PW2011} do not have any quantum corrections. As described above, this is due to the fact that the integral in \cite{PW2011} was not assumed to be of the form (\ref{Xquantparity}) but instead as in (\ref{Xenvelop})
\be \fl X=P_4(\hat{p}_1, \hat{p}_2, \hat{L}_3) +\frac12 \{ \tilde{f}_{2,2} , \hat{p}_1^2\} +\frac12 \{ \tilde{f}_{1,2} , \hat{p}_1p_2\} +\frac12 \{ \tilde{f}_{0,2} , \hat{p}_2^2\} +\tilde{f}_{0,4},\ee 
with
\bea\fl  \tilde{f}_{0,2}=f_{0,2}-\hbar^2\left(3y^2A_{400}-\frac32 A_{310}y+\frac12 c_1x^2+c_2x+c_3\right)\\
\fl \tilde{f}_{1,2}=f_{1,2}-\hbar^2\left(6yxA_{400}-c_1xy-3xA_{310}-2c_4x+\frac32 yA_{301} -c_2y+c_5\right)\\
\fl \tilde{f}_{2,2}=f_{2,2}-\hbar^2\left(3x^2A_{400}+\frac32A_{301}+\frac12 c_1y^2+c_4y+c_6\right),\eea
where the coefficients $c_j$ depend on the choice of symmetrization.

 For the final set of determining equations, there are two that need to be satisfied, instead of one as in the case $N=3$. They are given by 
\bea0=  2\partial_xf_{0,4}-\left(2f_{2, 2} \partial_xV+f_{1, 2} \partial_yV +\hbar^2 Q_{1,4}\right),\\
0=2\partial_yf_{0,4}-\left(f_{1, 2} \partial_xV+2f_{0, 2} \partial_yV +\hbar^2 Q_{0,4}\right),\eea
with 
\begin{eqnarray*}
\fl Q_{1,4}&=&-V_{yyy}f_{1,0}-2V_{xyy}f_{2,0}-3V_{xxy}f_{3,0}-4V_{xxx}f_{4,0}\\ \fl &&
-V_{yy}\left(-(6x^3+12xy)A_{400}+-6xyA_{310}-(\frac92x^2-3y^2)A_{301}-2yA_{211}-3xA_{202}+2xA_{220}-\frac32A_{103}+A_{121}\right)\\ \fl &&
-2V_{xy}\left((12x^2y-6y^3)A_{400}-(3x^2-\frac92y^2)A_{310}+6xyA_{301}+2yA_{202}-3yA_{220}-2xA_{211}+\frac32A_{130}-A_{112}\right)\\ \fl &&
-3V_{xx}\left(-6xyA_{400}+3xyA_{310}-\frac32A_{301}-xA_{202}+yA_{211}-\frac12A_{121}
\right)\\ \fl &&
-V_{y}\left(48xyA_{400}+12yA_{301}-12xA_{310}-4A_{211}\right)\\ \fl &&-2V_{x}\left((6x^2-18y^2)A_{400}+9yA_{310}+3xA_{301}+A_{202}-3A_{220}\right)\\ \fl &&
+ \partial_y^2\partial_xf_{0,2}+\frac32 \partial_y\partial_x^2 f_{1,2}+2\partial_x^3f_{2,2}+\frac12\partial_y^3f_{1,2}+\partial_y^2\partial_xf_{2,2}
\\
\fl Q_{0,4}&=&-4V_{yyy}f_{0,0}-3V_{xyy}f_{1,0}-2V_{xxy}f_{2,0}-V_{xxx}f_{3,0}\\ \fl &&
-3V_{yy}\left(-6x^2yA_{400}+\frac32x^2A_{310}-3xyA_{301}+xA_{211}-yA_{202}+\frac12A_{112}\right)\\ \fl &&
-2V_{xy}\left((6x^3+12xy^2)A_{400}-(\frac92x^2-3y^2)A_{301}-6xyA_{310}-3xA_{202}+2xA_{220}-2yA_{211}-\frac32A_{103}+A_{121}\right)\\ \fl &&
-V_{xx}\left((12x^2y-6y^3)A_{400}+6xyA_{301}-(3x^2-\frac92y^2)A_{310}+2yA_{202}-3yA_{220}-2xA_{211}+\frac32A_{130}-A_{112}
\right)\\ \fl &&
+2V_{y}\left((18x^2-6y^2)A_{400}+9xA_{301}+3yA_{310}+3A_{202}-A_{220}\right)\\ \fl &&-V_{x}\left(48xyA_{400}-12xA_{310}+12yA_{301}-4A_{211}\right)\\ \fl &&
+2 \partial_y^3f_{0,2}+\frac32 \partial_y^2\partial_x f_{1,2}+\partial_y\partial_x^2f_{2,2}+\partial_y\partial_x^2f_{0,2}+\frac12 \partial_x^3f_{1,2}
\end{eqnarray*}
Again, these determining equations could have quantum corrections quartic  in $\hbar$ but instead depend only on $\hbar^2.$ 
To compare these determining equations with those of \cite{PW2011}, we use the differential consequences of (\ref{f023}-\ref{f323}).

\subsection{Case $N=5$}
The integral we would like to investigate is 
\be  X=\frac12 \sum_{\ell=0}^{2}\sum_{j=0}^{5-2\ell} (-i\hbar)^{5-2\ell}\lbrace f_{j,2\ell}, \partial_x^j\partial_y^{5-2\ell-j}\rbrace.\ee
The highest-order determining equations are the same as in the all cases (\ref{enveloping}), with solutions \eref{fj0 Nthorder}.
However, unlike previous cases,   the next family of determining equations $M_{j,2}=0$ contain non-trivial quantum correction terms,  given by 
\be Q_{j,2} =\left(2\partial_x\phi_{j-1,2}+2\partial_y\phi_{j,2} +\partial_x^2\phi_{j,1}+\partial_y^2\phi_{j,1}\right).\ee
Unlike the lower-order cases, these quantum correction terms do not vanish and the determining equations are
\begin{eqnarray*}
\fl \partial_yf_{0,2}&=&\frac12 f_{1,0}\partial_xV+\frac52 f_{0,0}\partial_yV   
 						    	-\hbar^2\left(\frac32A_{{311}}+6A_{{410}}x-30A_{{500}}xy-6A_{{401}}y\right)\\\fl
\partial_xf_{0,2}+\partial_yf_{1,2}&=& f_{2,0}\partial_xV+2 f_{1,0}\partial_yV						\\	\fl											
&&-\hbar^2\left( 30\,A_{{500}}{y}^{2}-
30\,A_{{500}}{x}^{2}-12\,A_{{410}}y-12\,A_{{401}}x+3\,A_{{320}}-3\,A_{{302}}
\right)\nn\fl
\partial_xf_{1,2}+\partial_yf_{2,2}&=& \frac32 f_{3,0}\partial_xV+\frac32 f_{2,0}\partial_yV,\nn\fl
\partial_xf_{2,2}+\partial_yf_{3,2}&=&  2f_{4,0}\partial_xV+f_{3,0}\partial_yV\nn\fl
&&				-\hbar^2\left( 30\,A_{{500}}{y}^{2}-
30\,A_{{500}}{x}^{2}-12\,A_{{410}}y-12\,A_{{401}}x+3\,A_{{320}}-3\,A_{{302}}
\right),\nn\fl
\partial_xf_{3,2} &=&  \frac52f_{5,0}\partial_xV+\frac12 f_{4,0}\partial_yV
					-\hbar^2\left(6A_{{401}}y+30A_{{500}}xy-\frac32A_{{311}}-6A_{{410}}x\right).
				\end{eqnarray*}
			Note that while these equations contain possibly non-trivial quantum corrections, they can be removed by taking the form of $X$ as in (\ref{Xenvelop}).
We omit the remainder of the equations as they quickly become unmanageable, although it is interesting to note that in both $M_{j,4}$ and $M_{0,6}$ the highest degree of $\hbar$ is 2 and 4 respectively; that is,  as in the cases for $N\leq3,$ the highest-order terms in $\hbar$ are missing from the two lowest families of equations.

 \section{Quantizing Classical Operators}
 In this section, we would like to make a few comments and observations about the relationship between classical and quantum integrals. It is clear from the previous sections that, given a quantum integral of motion of the form in \eref{Xquantparity}
 that commutes with a Hamiltonian $H$, and assuming that both the  potential $V$ and the functions $f_{j,2\ell}$ are independent of $\hbar$, then the classical integral 
 \be \mathcal{X}= \sum_{\ell=0}^{[\frac{N}2]}\sum_{j=0}^{N-2\ell} f_{j,2\ell}, p_1^jp_2^{N-2\ell-j},\ee
 will be an integral of the motion for a classical Hamiltonian with the same potential. 
 
 However, the implication is clearly not reversible. This observation and the general question of quantizing a classical system has received much interest over the years,  see for example \cite{Hiet1984, HG1989, hietarinta1998pure}. One of the first questions that appear in the transition from the classical to the quantum case is the choice of symmetrization, assuming cannonical quantization. Clearly, the symmetrization used in  (\ref{Xquantparity}) is not unique. Indeed, there are many possible choices. The most general choice would be 
 \be f_{j,2\ell}p_1^jp_2^{N-2\ell-j}\rightarrow S(f_{j,2\ell}, c_{a,b})\label{generalsymm}\ee
 with 
 \[ \fl S(f_{j,2\ell}, c_{a,b})\equiv \sum_{b=0}^{N-2\ell}\sum_{a=0}^{j}c_{a,b}\left(\hat{p}_1^a\hat{p}_2^{b-a}f_{j,2\ell}\hat{p}_1^{j-a}\hat{p}_2^{N-2\ell-j-b+a}+\hat{p}_1^{j-a}\hat{p}_2^{N-2\ell-j-b+a}f_{j,2\ell}\hat{p}_1^a\hat{p}_2^{b-a}\right),\]
 with $\sum_{a,b}c_{a,b}=\frac12$. 
 It is true, however, that the choice of symmetrization does not affect the general form of the integral and furthermore the different choices will lead to quantum correction terms in the $f_{j,2\ell}$ as polynomials in $\hbar^2.$ In particular, given  a choice of symmetrization as in 
 (\ref{generalsymm}),  we shall show that this choice is equivalent to the standard one chosen in (\ref{Xquantparity}) with quantum corrections to the coefficient functions. The determining equations for any choice of symmetrization in (\ref{generalsymm}) are equivalent to
the standard one chosen in (\ref{Xquantparity}) up to appropriate modifications in the
quantum corrections $Q_{j,2\ell}$. The formulas given in Theorem 2 only hold for
for the canonical choice (\ref{Xquantparity}).
 
 \begin{theorem} Let $f_{j,2\ell}\in C^{\infty}(\mathbb{R}^2)$ be  polynomial in $\hbar^2$, then 
 the self-adjoint differential operator  $S(f_{j,2\ell}, c_{a,b})$ defined in (\ref{generalsymm}) can be expressed as 
\bea \fl S(f_{j,2\ell}, c_{a,b})&=&\frac12\lbrace f_{j,2\ell}, \hat{p}_1^{j}\hat{p}_2^{N-2\ell-j}\rbrace\label{symhypoth}-\hbar^2\sum_{k=0}^{[N-2\ell]}\left(\frac12\lbrace g_{j,2\ell+2k}, \hat{p}_1^{j}\hat{p}_2^{N-2\ell-2k-j}\rbrace \right),\eea
where the $g_{j,2\ell+2k}$ are polynomial in $\hbar^2.$
 \end{theorem}
By Lemma \ref{sym}, the differential operator $ S(f_{j,2\ell}, c_{a,b})$ can be expressed as 
the sum 
\be   S(f_{j,2\ell},  c_{a,b})=\frac12\sum_m\sum_n \lbrace g_{n,2\ell+m}, \hat{p}_1^n\hat{p}_2^{N-2\ell-m-n}\rbrace.\ee
Because of the requirement $\sum c_{a,b}=\frac12$, the leading order terms will agree, $g_{j,2\ell}=f_{j,2\ell}$. The next order term will have a different parity under time reversal (complex conjugation) and so will vanish identically leaving a differential operator of degree 2 less multiplied by a factor of $-\hbar^2$. Thus, by induction, we see that (\ref{symhypoth}) holds. 

Note that the case of quantizing classical integrals would correspond to the functions $f_{j,2\ell}$ being independent of $\hbar^2$, in the theorem. In what follows we give an example of a classical system where the choice of symmetrization is slightly non-intuitive. 
 
 \subsection{Example I}
 Let us consider an example of a potential that allows separation of
variables in polar coordinate\cite{TW20101}, has a nonzero $\hat{p}_2(\hat{L}_3)^2$ term, and  has a non-zero classical limit. This example is in fact second order superintegrable and the third order integral given below can be obtained from the lower order integrals. 
The potential is 
\be V=\frac{a}{\sqrt{x^2+y^2}}+\frac{\alpha_1}{x^2}+\frac{\alpha_2y}{x^2\sqrt{x^2+y^2}}\ee
and it allows separation of variables in polar and parabolic coordinates. 
The classical third-order integral is given by 
\be \mathcal{X}=p_2L_3^2+f_{1,2}p_1+f_{0,2}p_2\ee
with 
\[ f_{1,2}=-\frac{x(ay-\alpha_2)}{2\sqrt{x^2+y^2}}\]
\[ f_{0,2}=\frac{ax^2+2\alpha_2y}{2\sqrt{x^2+y^2}}+\frac{\alpha_2y^3}{x^2\sqrt{x^2+y^2}}+\frac{\alpha_1(x^2+y^2)}{x^2}.\] 
The correct symmetrization which keeps $f_{1,2}$ and $f_{0,2}$ fixed is given by 
\be X=\frac{1}{8}(\hat{p}_2\hat{L}_3^2+2\hat{L}_3 \hat{p}_2\hat{L}_3+\hat{L}_3^2\hat{p}_2)+\frac{1}{2}\lbrace f_{1,2}, \hat{p}_1\rbrace+ \frac{1}{2}\lbrace f_{0,2}, \hat{p}_2\rbrace.\label{Lparticular}\ee
 This integral can be expressed in the standard form (\ref{Xquantparity}) via 
 \[ f_{0,0}=x^2, \qquad f_{1,0}=-2xy, \qquad f_{2,0}=y^2, \qquad f_{3,0}=0\]
 and a quantum correction to $f_{0,2}$ of $7/4\hbar^2$ so that (\ref{Lparticular}) becomes
 \be \fl X=\sum_{j=0}^{3}\frac{1}{2}\lbrace f_{j,0}, \hat{p}_1^{j}\hat{p}_2^{3-j}\rbrace+\frac{1}{2}\lbrace f_{1,2}, \hat{p}_1\rbrace+\frac{1}{2}\lbrace f_{0,2}+\frac74 \hbar^2, \hat{p}_2\rbrace.\ee
 Note that the appropriate symmetrization (keeping the lower order-terms free of $\hbar$ dependent terms) is neither $\{L_3^2,p_2\}$ as in\cite{Gravel, TW20101} nor simply $\lbrace f_{j,0}, \partial_x^{j}\partial_y^{3-j}\rbrace$ as in \eref{Xquantparity}.

 \subsection{Example II}
 In this section, we consider quantum Hamiltonian systems associated with exceptional Hermite polynomials of Marquette and Quesne \cite{marquette2013new}. We mention also that these systems are composed of 1D exactly solvable Hamiltonians for exceptional Hermite polynomials discovered and analyzed by G\'omez-Ullate, Grandati, and Milson \cite{gomez2014rational}. 
 
 The first example of a rational extension of a simple Harmonic oscillator is given by 
 \be \label{Hex1} H=-\hbar^2\left(\partial_x^2+\partial_y^2\right)+\omega^2(x^2+y^2)+\frac{8\hbar^2\omega(2\omega x^2-\hbar)}{(2\omega x^2+\hbar)^2}.\ee
The Hamiltonian admits separation of variables in Cartesian coordinates and  hence a second-order integral of the motion 
\be \label{X1} X_1= \hat{p}_2^2+\omega^2y^2.\ee
Additionally, there are two, third-order integral of the motion given by 
\begin{eqnarray}\fl  X_2&=&\frac12\{ {L}_3, \hat{p}_1 \hat{p}_2\}+ \{\omega^2x^2,  \hat{L}_3\}\\\fl
&&+\hbar\left(\{2\omega y +\frac{12\hbar \omega y(2\omega x^2-\hbar)}{(2\omega x^2+\hbar)^2},  \hat{p}_1\}-\{2\omega x +\frac{12\hbar \omega y(2\omega x^2-\hbar)}{(2\omega x^2+\hbar)^2},  \hat{p}_2\}\right),\nonumber \eea
\bea\fl X_3&=& \hat{L}_3^3+\frac{3\hbar}{2\omega}\left[ \frac12\{ \hat{L}_3,X_1\}\right. \\ \fl &&
\qquad \qquad +\left.\hbar\left(2 {L_3}-\lbrace\frac{8\hbar\omega^2y^3(2\omega x^2+\hbar)}{(2\omega x^2+\hbar)^2},  \hat{p_1}\rbrace+\lbrace\frac{8\hbar\omega^2xy^2(2\omega x^2+3\hbar)}{(2\omega x^2+\hbar)^2},  \hat{p_2}\rbrace
\right)\right].\nonumber \end{eqnarray}
This system can thus be considered as a quantum deformation of the harmonic oscillator. As demonstrated by the existence of the third-order integrals, this system falls in the classification of Gravel \cite{Gravel}. The connection can be directly obtained by setting $\omega=-\hbar/(2\alpha)^2$ to obtain  the potential $V_{e}$ in the classification\cite{Gravel}.


Another example is based on the fourth Hermite polynomial. The associated superintegrable Hamiltonian is given by 
 \[ \fl H=-\hbar^2\left(\partial_x^2+\partial_y^2\right)+\omega^2(x^2+y^2)+{\frac {768{\hbar}^{4}{\omega}^{2}{x}^{2}
}{ \left( 4\,{\omega}^{2}{x}^{4}+12\,\omega\,{x}^{2}\hbar+3\,{h}^{2}
 \right) ^{2}}}+\,{\frac {16{\hbar}^{2}\omega\, \left( 2\,\omega\,{x}^{2}
-3\,\hbar \right) }{4\,{\omega}^{2}{x}^{4}+12\,\omega\,{x}^{2}\hbar+3\,{\hbar}^{2}
}}.
\]
The Hamiltonian admits $X_1,$ as in (\ref{X1}), as well as the following third-order integral
\bea\fl  X_2&=&\{ \hat{L}_3,  \hat{p}_1^2\}+\{\omega^2x^2,  \hat{L}_3\}\nn\fl
&&+\hbar\left[6\omega  \hat{L}_3-\left\{\frac{24\hbar\omega y(8\omega^2x^6+12\hbar \omega^2x^4+18\hbar^2\omega x^2-9\hbar^3)}{4\omega^2x^4+12\hbar \omega x^2+3\hbar^2)^2},  \hat{p}_1\right\}\right.\nn
\fl &&\qquad \qquad\quad  \left.+\left\{\frac{8\hbar\omega y(8\omega^2x^6-12\hbar \omega^2x^4-6\hbar^2\omega x^2-27\hbar^3)}{4\omega^2x^4+12\hbar \omega x^2+3\hbar^2)^2},  \hat{p}_2\right\}\right].
\eea
Unlike the previous case, this potential is not immediately recognizable in Gravel's classification. A remarkable fact is that this potential is associated with the fourth Painlev\'e equation. Indeed, the $x$-dependent part of the potential $W(x)=V(x,y) -\omega^2y^2+4\omega\hbar$ satisfies 
\[-\hbar^2W^{(4)}-12\omega^2(xW)'+3(W^2)''-2\omega^2x^2W''+4\omega^4x^2=0,\]
which is equivalent to Eq. (17) of the paper \cite{Gravel} by scaling the potential and the variable. It can also be shown that the potential of the  previous Hamiltonian (\ref{Hex1}) is also a solution of this non-linear equation. Thus, these two systems whose wave functions are given by exceptional Hermite polynomials have potentials that can be expressed in terms of rational solutions to the fourth Painlev\'e equation\cite{Gromak1999, GLS}. Of course, many such particular solutions to the Painlev\'e equations exist but their connection to exceptional orthogonal polynomials as well as the harmonic oscillator is quite remarkable and will be investigated in future work.

\section{Conclusions}
The main results of this article are summed up in Theorems 1 and 2. They present the determining equations for the coefficients of an Nth order integral of the motion $X$ in the Euclidean plane $E_2$ in classical and quantum mechanics, respectively. Both the similarities and differences between the two cases are striking. 

The number of determining equations $M_{j, 2\ell}=0$ to solve and the number of coefficient functions $F_{j, 2\ell}(x,y)$ to determine is the same in the classical and quantum cases (see (\ref{numdet}) and (\ref{numcof})). If the potential $V(x,y)$ is known, the determining equations are linear. If the potential is not known a priori then in both cases we have a coupled system of nonlinear PDE for the potential $V(x,y)$ and the coefficients $f_{j, 2\ell}$ ($0\leq \ell \leq \lfloor \frac{N}{2}\rfloor$ $0\leq j\leq N-2\ell$). In both cases the functions $f_{j, 2\ell}$ are real, if the potential is real and the quantum integral is assumed to be a Hermitian operator.   In both cases the integral $X$ contains only terms of the same parity as the leading terms (obtained for $\ell=0$). The leading terms in the polynomial lie in the enveloping algebra of the Euclidean Lie algebra. This is a consequence of the fact that the determining equations $M_{j, 2\ell}=0$ for $\ell=0$ do not depend on the potential and are the same in both the classical and quantum case. For $N\geq 3$ the quantum determining equations with $\ell\geq 1$ have quantum corrections. For $\ell=1$ the determining equations $M_{j, 2}=0$ for $j$ in the interval $0\leq j\leq N$ must satisfy a compatibility condition. This linear compatibility condition (\ref{linearcomp}) for the potential to allow and Nth order integral is the same in the classical and quantum case. However, if we consider the determining equations (\ref{classical deteq}) and (\ref{quant deteq})  respectively for $\ell\geq 2$ they will differ in the two cases. Moreover, new compatibility conditions on the potential arise for each higher value of $\ell$. They will be nonlinear equations for $V(x,y)$ and will be considerably more complicated in the quantum case than in the classical one (see \cite{ Gravel, GW, MPSnonlin2014, MW2008, TW20101, PopperPostWint2012} for the case $N=3$ and Section 3 of this article for $N\geq 3$). 

As mentioned above, the number of determining equations to solve for any given $N$ is actually smaller than given by (\ref{numdet}) since those for $f_{j,0}$ have been solved for general $N$ (there are $N+2$ such equations). Even so, solving the determining equations for $N>2$ is a formidable task, even for $N=3$. A much more manageable task is to use the determining equations in the context of superintegrability. In a two-dimensional space a Hamiltonian system is superintegrable if it allows two integrals of motion $X$ and $Y$, in addition to the Hamiltonian. They satisfy $[X,H]=[Y, H]=0$. Assuming that $X$ and $Y$ are polynomials in the momenta and that the system considered is defined on $E_2$, both will have the  form studied in this article. The integrals $H$, $X$ and $Y$ are assumed to be polynomially independent. The integrals $X$ and $Y$ do not commute, $[X,Y]\ne 0,$ and hence generate a non-Abelian polynomial algebra of integrals of the motion. 

The case that has recently been the subject of much investigation is that when one of the integrals is of order one or two and hence the potential will have a specific form that allows separation of variables in the Hamilton-Jacobi and Schr\"odinger equations. The potential $V(x,y)$ will then be written in terms of two functions of one variable each, the variables being either Cartesian, polar, parabolic or elliptic coordinates. Once such a potential is inserted into the determining equations, they become much more manageable. 

For $N=3$ and $N=4$ \cite{Gravel, GW,  MPSnonlin2014, MW2008,  TW20101, PopperPostWint2012, MSW2015} this procedure, at least for Cartesian and polar coordinates leads to new superintegrable systems, not obtained when both integrals $X$ and $Y$ are of order $N\leq 2.$ In classical mechanics these potentials are expressed in terms of elementary functions or solutions of algebraic equations\cite{GW, MW2008, TW20101}. In quantum mechanics one also obtains ``exotic potentials" that do not satisfy any linear PDE, i.e. the linear compatibility condition (\ref{linearcomp}) is solved trivially, see \cite{MPSnonlin2014}. These exotic potentials are expressed in terms of elliptic functions  or Painlev\'e transcendents \cite{Gravel, GW, MW2008, TW20101}. 

This has so far been done systematically for $N=3$ and $N=4$. The formalism presented in this article makes it possible to investigate superintegrable separable potentials for all $N$. 
Another application of the presented formalism is to make assumptions about the form of the potential and then look for possible integrals of the motion. Hypotheses about integrability or superintegrability of a given potential can then be verified (or refuted) by solving a system of linear PDEs. 

Alternative approaches to the construction of superintegrable systems in two or more dimensions exist. In quantum mechanics they typically start from a one-dimensional Hamiltionian $H_1=p_1^2+V_1(x)$  and a Hermitian operator $X_1$ that satisfies some predetermined commutation relation with the Hamiltonian. A very recent article \cite{GKNN} was devoted to the ``Heisenberg symmetries," when the commutation relation was chosen to be $[H_1, X_1]=1$. Combined with a similar relation in the $y$ coordinate $[H_2, X_2]=1,$ this leads to interesting superintegrable systems. For other assumptions, e.g. that $X_1$ is a one-dimensional ladder operator, see \cite{Marquette2012, marquette2013new} and references therein.

There are many avenues open for further research. In addition to the already discussed applications to Nth order superintegrability already in progress, we mention the extension of the theory to spaces of non-zero curvature \cite{ballesteros2011quantum, ballesteros2014exactly, BEHRR} and to higher dimensions \cite{riglioni2013classical, PostRiglioni2014}. 

\ack S.P. thanks the Centre de Recherches Math\'ematiques for their hospitality over several visits to complete this research.  The research of PW was partially supported by NSERC of Canada. He thanks the European Union Research Executive Agency for the award of a Marie Curie International Incoming Research Fellowship making his stay at Universit\`a Roma Tre possible. He thanks the Department of Mathematics and
Physics of the Universit\`a Roma Tre  and especially Professor Decio Levi for hospitality.

\section*{References}
\bibliography{bib}{}
\bibliographystyle{plain}
\end{document}